\documentclass[a4paper,11pt]{article}
\usepackage{pos}

\title{Heavy-quark dynamics in QGP using effective field theory}
\ShortTitle{Heavy-quark dynamics in QGP using effective field theory}

\author[a]{Andreas Kirchner}

\author[a]{Berndt Müller}

\author*[a]{Jyotirmoy Roy}

\author[a]{Chathuranga Sirimanna}

\affiliation[a]{Department of Physics, Duke University,\\
  Durham, NC 27708, USA}

\emailAdd{andreas.kirchner@duke.edu, berndt.mueller@phy.duke.edu, jyotirmoy.roy@duke.edu, chathuranga.sirimanna@duke.edu}

\abstract{Heavy quarks are sensitive to the bulk properties of the quark-gluon plasma, yet the soft exchange of energy and momentum between a heavy quark and the plasma cannot be computed reliably in perturbation theory. We formulate an effective field theory in which the plasma is represented by three scalar fields that label its fluid elements, carry no color and interact only through their derivatives, and we couple these fields to a heavy quark and to gluons. Symmetries and a derivative expansion fix the form of the interactions, and power counting in the hydrodynamic UV scale organizes them systematically. The fluctuations of the fluid are phonons, the weakly coupled collective excitations of the medium. Within this framework we compute the quasiparticle width of the heavy quark, which grows rapidly once the quark moves faster than sound because phonon emission becomes kinematically allowed, and the heavy quark--phonon scattering cross section, which is resonantly enhanced when the Mach condition is satisfied.}

\FullConference{13th International Conference on Hard and Electromagnetic Probes of High-Energy Nuclear Collisions (HardProbes2026)\\
21-26 June 2026\\
Nashville, Tennessee, USA\\}

\usepackage{subcaption}
\usepackage{wrapfig}

\begin{document}
\maketitle

\section{Introduction}

Hard probes such as heavy quarks and jets originate in the initial hard scatterings of a heavy-ion collision and traverse the entire evolution of the quark-gluon plasma (QGP), which makes them sensitive to its bulk and transport properties. They also leave an imprint on the medium: CMS has observed the diffusion wake of jets in dijet events~\cite{CMS:2026wake}. Such wakes are driven by the interaction of the probe with the collective modes of the fluid, and holographic calculations of a heavy quark moving through a plasma display both a Mach cone and a diffusion wake~\cite{Gubser:2007ga,Chesler:2007sv}.

Describing this interaction from first principles is difficult: at the temperatures reached in heavy-ion collisions the QCD coupling is not small, and perturbative expansions converge poorly; the next-to-leading-order correction to the heavy-quark momentum diffusion coefficient, for example, is large for realistic couplings~\cite{Caron-Huot:2007rwy}. Moreover, at soft momenta the relevant excitations of the medium are collective modes, which have no simple description in terms of partons. What is needed is a framework formulated directly in terms of these modes that incorporates thermal and quantum fluctuations systematically, and effective field theory (EFT) meets both requirements. In Ref.~\cite{Kirchner:2025ehq} we constructed such an EFT, the heavy quark+hydro EFT (QHEFT), which describes the dynamics of heavy quarks coupled to the hydrodynamic modes of the medium. Here we summarize it and present the heavy-quark width and the heavy quark--phonon cross section.

\section{The heavy quark+hydro effective field theory}

The QHEFT has three types of degrees of freedom: the hydrodynamic fields $\phi^I(x)$, $I=1,2,3$, the heavy-quark field $\psi(x)$, and the gluon field $A^a_\mu(x)$. Following the EFT description of perfect fluids~\cite{Dubovsky:2005xd,Endlich:2010hf,Dubovsky:2011sj}, $\phi^I(x)$ are the comoving (Lagrangian) coordinates of the fluid elements, with $\phi^I = x^I$ in equilibrium. Homogeneity, isotropy and the absence of resistance to shape deformations at fixed volume (non-elasticity) of the ground state are imposed as the internal symmetries
\begin{equation}
\text{(i)}\ \phi^I \to \phi^I + \alpha^I, \qquad
\text{(ii)}\ \phi^I \to R^I{}_J\,\phi^J,\ \ R\in SO(3), \qquad
\text{(iii)}\ \phi^I \to \xi^I(\phi),\ \ \det\!\left(\frac{\partial \xi^I}{\partial \phi^J}\right)=1 ,
\label{eq:sym}
\end{equation}
with constant $\alpha^I$; symmetry (iii) allows any volume-preserving relabeling of the fluid elements and has no counterpart for an elastic solid. With one derivative acting on each field, the only invariant built from $\phi^I$ at the lowest order is the current
\begin{equation}
J^\mu = \frac{1}{3!}\,\epsilon^{\mu\alpha\beta\gamma}\,\epsilon_{IJK}\,\partial_\alpha\phi^I\,\partial_\beta\phi^J\,\partial_\gamma\phi^K ,
\qquad X \equiv J^\mu J_\mu ,
\label{eq:J}
\end{equation}
which is invariant under (i)--(iii) and identically conserved; with the metric $\mathrm{diag}(+,-,-,-)$, $X=1$ for the fluid at rest.

We couple $\phi^I$ to the heavy quark and the gluons through $J^\mu$. Since $\partial_\mu\phi^I$ is dimensionless, so are $J^\mu$ and $X$, and functions of $X$ do not change the dimension of an operator. The lowest order of the derivative expansion consists of the relevant and marginal operators~\cite{Kirchner:2025ehq},
\begin{align}
\mathcal{L} = {}& F(X) + Z_\psi(X)\,\bar\psi\,\tfrac{i}{2}\gamma^\mu \overleftrightarrow{D}_{\!\mu}\psi - M(X)\,\bar\psi\psi + Z_A(X)\,G^a_{\mu\nu}G_a^{\mu\nu} \nonumber\\
& + P(X)\,\bar\psi\gamma_\mu\psi\,J^\mu + Q(X)\,\bar\psi\,\tfrac{i}{2}\overleftrightarrow{D}_{\!\mu}\psi\,J^\mu + R(X)\,\bar\psi\sigma_{\mu\nu}\psi\,\omega^{\mu\nu} .
\label{eq:L}
\end{align}
Here $D_\mu = \partial_\mu - i g A^a_\mu t^a$ is the covariant derivative, $\overleftrightarrow{D}_{\!\mu} = \overrightarrow{D}_{\!\mu}-\overleftarrow{D}_{\!\mu}$, $G^a_{\mu\nu} = \partial_\mu A^a_\nu - \partial_\nu A^a_\mu + g f^{abc}A^b_\mu A^c_\nu$ is the gluon field strength, $\sigma_{\mu\nu}=\tfrac{i}{2}[\gamma_\mu,\gamma_\nu]$, and $\omega_{\mu\nu}=\partial_\mu J_\nu - \partial_\nu J_\mu$ is the vorticity of the fluid current. Because the fluid is colorless, $J^\mu$ enters only in color-singlet combinations. The coefficients $F, Z_\psi, M, Z_A, P, Q, R$ are generic functions of $X$ that play the role of Wilson coefficients: they encode the nonperturbative physics of the medium and must be determined by matching or from data. Operators with more derivatives are irrelevant and suppressed by a scale $\Lambda_{\rm mfp}$ at which the fluid description itself breaks down, so Eq.~\eqref{eq:L} holds for $\partial_\mu\ll\Lambda_{\rm mfp}$; only $M(X)$ enters the results below~\cite{Kirchner:2025ehq}.

The pure hydrodynamic term $F(X)$ describes a perfect fluid, with energy-momentum tensor
\begin{equation}
T_{\mu\nu} = (\varepsilon+P)\,u_\mu u_\nu - P\,g_{\mu\nu}, \qquad
\varepsilon = -F(X), \qquad P = F(X) - 2XF'(X), \qquad u^\mu = \frac{J^\mu}{\sqrt{X}} ,
\label{eq:Tmunu}
\end{equation}
so $\varepsilon$ and $P$ depend on $X$ alone, and $F(X)$ encodes the equation of state $P=P(\varepsilon)$, consistent with the thermodynamic identities. For example, $F(X) = -\varepsilon_0\,X^{(1+c_s^2)/2}$ gives $P = c_s^2\,\varepsilon$, a fluid with a constant speed of sound $c_s$. At this order the fluid is ideal; dissipative effects enter through corrections to the leading-order theory~\cite{Endlich:2012vt,Grozdanov:2013dba}.

\begin{figure}[t]
\centering
\includegraphics[width=0.75\textwidth]{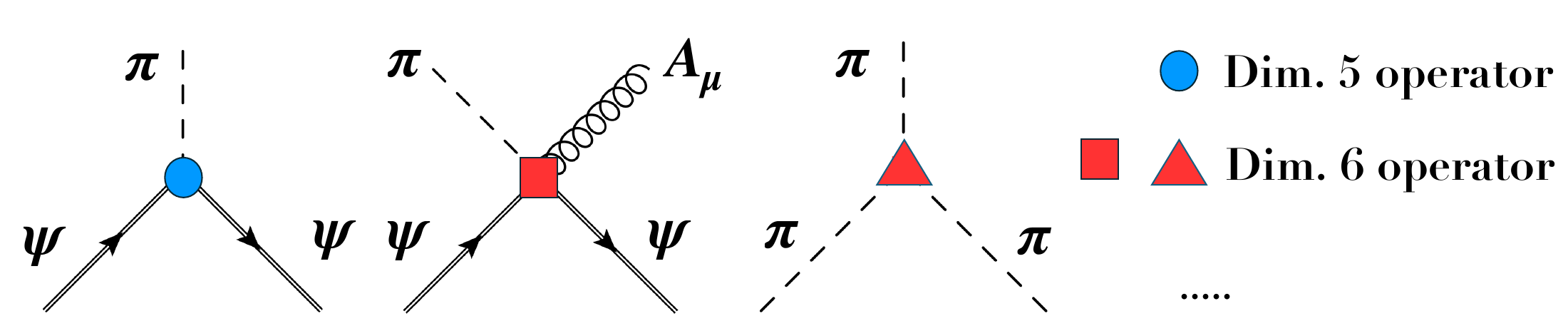}
\caption{Leading QHEFT vertices: heavy quark--phonon (dimension five), heavy quark--phonon--gluon and three-phonon (dimension six).}
\label{fig:vertices}
\end{figure}

\section{Fluctuations about equilibrium and power counting}

The excitations of the fluid are introduced by expanding about equilibrium, $\phi^I(x) = x^I + \pi^I(x)/\Lambda_h^2$. Expanding $F(X)$ to second order in $\pi^I$ gives the phonon Lagrangian density
\begin{equation}
\mathcal{L}_\pi = \frac{1}{2}\big(\partial_t \vec\pi\big)^2 - \frac{1}{2}\,c_s^2\big(\vec\nabla\cdot\vec\pi\big)^2 + \dots ,
\qquad
c_s^2 = \frac{2F''(X_0)X_0 + F'(X_0)}{F'(X_0)} ,
\label{eq:Lpi}
\end{equation}
where $X_0=1$ is the equilibrium value and the dots denote interactions. A canonical kinetic term fixes a second scale, the hydrodynamic UV scale $\Lambda_h$, through $\Lambda_h^4 \equiv -2F'(1)$, with $\Lambda_h<\Lambda_{\rm mfp}$~\cite{Kirchner:2025ehq}. The longitudinal component of $\vec\pi$ is the sound mode, $\omega_k = c_s|\vec k|$, whose quanta are the phonons; without viscosity the two transverse polarizations, the shear modes, do not propagate ($\omega=0$).

The phonons are the Goldstone modes of the symmetries spontaneously broken by the equilibrium configuration $\phi^I = x^I$, so they couple only derivatively. The higher-order terms in the expansion of $F(X)$ in $\pi^I$ are phonon self-interactions, and $\Lambda_h$ is the scale that suppresses them: the three-phonon term is a dimension-six operator suppressed by $1/\Lambda_h^2$, and an $n$-phonon vertex is suppressed by $\Lambda_h^{-(2n-4)}$~\cite{Kirchner:2025ehq}. The phonons are therefore weakly coupled for $|\vec k|\ll\Lambda_h$, whatever the coupling of the underlying medium, and can be treated perturbatively in $|\vec k|/\Lambda_h$. Requiring that sound waves be only weakly attenuated at the UV scale gives the estimate $3T \lesssim \Lambda_h \lesssim 6T$~\cite{Kirchner:2025ehq}.

The same expansion applied to the heavy-quark sector of Eq.~\eqref{eq:L} generates the phonon--heavy quark interactions shown in Fig.~\ref{fig:vertices}. To first order in $\pi^I$, $J^0 = \det(\partial_j\phi^i) \simeq 1 + \vec\nabla\cdot\vec\pi/\Lambda_h^2$ and $\vec J = \mathcal{O}(\partial_t\vec\pi)$, so $X \simeq 1 + 2\,\vec\nabla\cdot\vec\pi/\Lambda_h^2$. The mass term then gives
\begin{equation}
M(X)\,\bar\psi\psi = m\,\bar\psi\psi + \frac{2\tilde M'(1)}{\Lambda_h}\,\bar\psi\psi\;\vec\nabla\cdot\vec\pi + \mathcal{O}\!\left(\Lambda_h^{-3}\right),
\qquad m\equiv M(1), \qquad \tilde M'(1)\equiv \frac{M'(1)}{\Lambda_h},
\label{eq:vertex}
\end{equation}
with $\tilde M'(1)$ a dimensionless Wilson coefficient of natural size. The heavy quark thus couples to the compressional mode through a dimension-five operator. The heavy quark--phonon--gluon vertex, from the $X$ dependence of the covariant-derivative terms, is of dimension six, and two-phonon contact terms are suppressed by $\Lambda_h^{-3}$, so at leading order the dynamics is governed by Eq.~\eqref{eq:vertex}. Below we set $\tilde M'(1)=1$; the width and cross section scale as $\tilde M'(1)^2$.

\section{Quasiparticle width of the heavy quark}

Because the phonon dispersion $\omega_k = c_s|\vec k|$ lies below the light cone, a heavy quark can absorb and emit on-shell phonons. The kinematics of both processes follows from the on-shell conditions $(p\pm k)^2 = m^2$ for a heavy quark with momentum $p^\mu = (E_p,\vec p)$, velocity $v = |\vec p|/E_p$, and a phonon $k^\mu = (c_s|\vec k|,\vec k)$ at angle $\theta_k$ to $\vec p$:
\begin{equation}
\cos\theta_k = \frac{c_s}{v} \mp \frac{(1-c_s^2)\,|\vec k|}{2|\vec p|} ,
\label{eq:kin}
\end{equation}
where the upper sign corresponds to absorption, $Q(p)+\pi(k)\to Q(p+k)$, and the lower sign to emission, $Q(p)\to Q(p-k)+\pi(k)$. Since $|\cos\theta_k|\le 1$, emission (lower sign) requires $v>c_s$, the acoustic analog of Cherenkov radiation; the emitted momentum is then bounded by $|\vec k| \le 2|\vec p|(1-c_s/v)/(1-c_s^2)$. Absorption, in contrast, is possible at any velocity.

\begin{wrapfigure}{r}{0.47\textwidth}
\vspace{-14pt}
\centering
\includegraphics[width=0.46\textwidth]{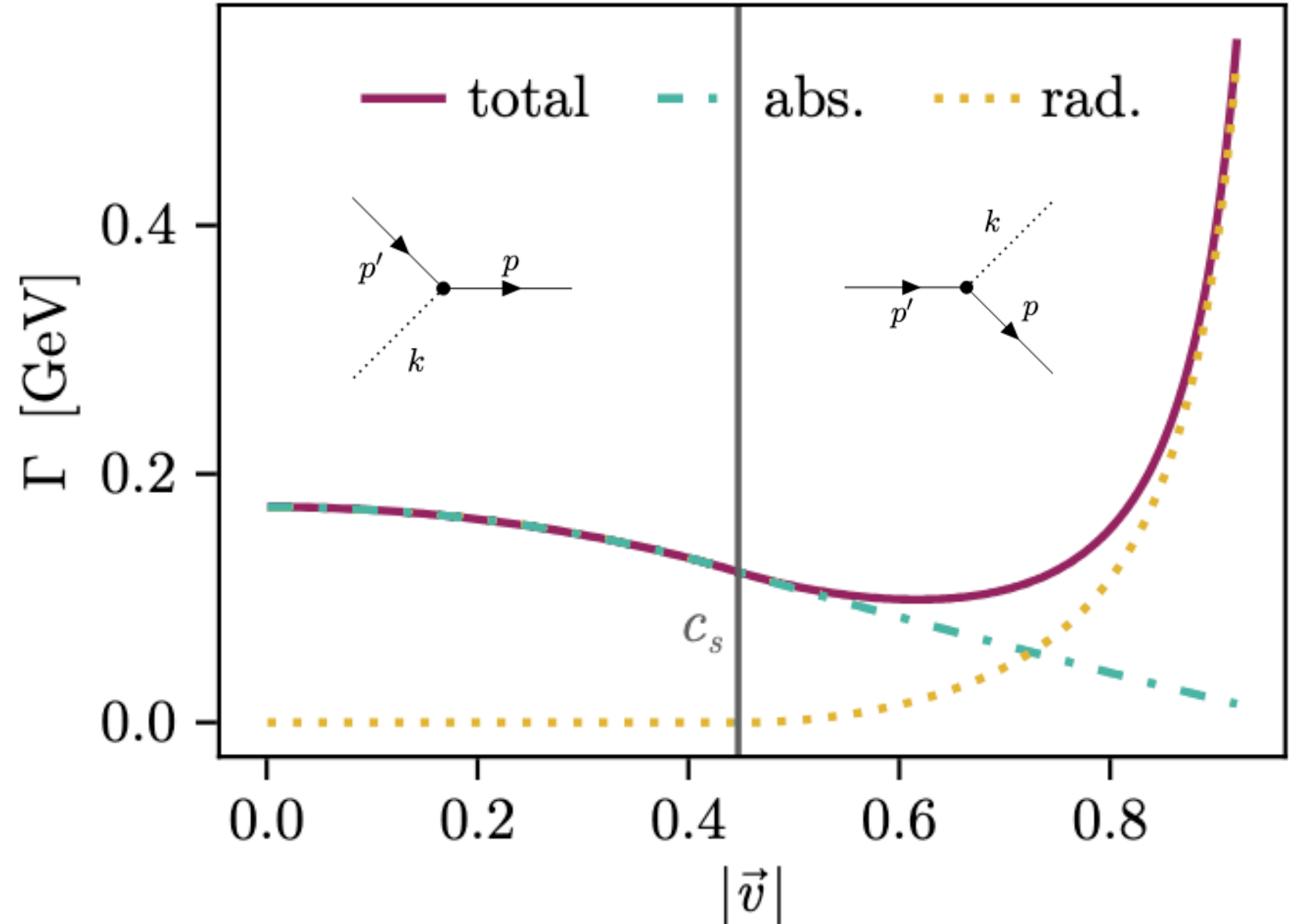}
\caption{Quasiparticle width of a charm quark vs.\ its speed $|\vec v|$ (solid), with absorption (dash-dotted) and radiation (dotted) contributions; the vertical line marks $c_s$. Adapted from Ref.~\cite{Kirchner:2025ehq}.}
\label{fig:width}
\vspace{-10pt}
\end{wrapfigure}
The heavy quark therefore acquires a finite quasiparticle width $\Gamma$ in the fluid, the sum of an absorption contribution weighted by the Bose distribution $n(\omega_k) = 1/(e^{\omega_k/T}-1)$ of thermal phonons and an emission contribution weighted by $1+n(\omega_k)$, which includes spontaneous and stimulated emission~\cite{Kirchner:2025ehq}. Both are proportional to $\tilde M'(1)^2/\Lambda_h^2$ from the vertex of Eq.~\eqref{eq:vertex}. Figure~\ref{fig:width} shows $\Gamma$ for a charm quark. Below the speed of sound the width comes entirely from absorption and decreases slowly with $v$. Once $v>c_s$ the emission channel opens and soon dominates, so that $\Gamma$ grows steeply with $v$.

\begin{figure}[t]
\centering
\begin{subfigure}[b]{0.36\textwidth}
\centering
\includegraphics[height=4.7cm]{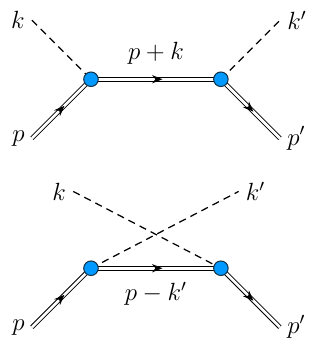}
\caption{}
\label{fig:diagrams}
\end{subfigure}\hfill
\begin{subfigure}[b]{0.6\textwidth}
\centering
\includegraphics[height=4.7cm]{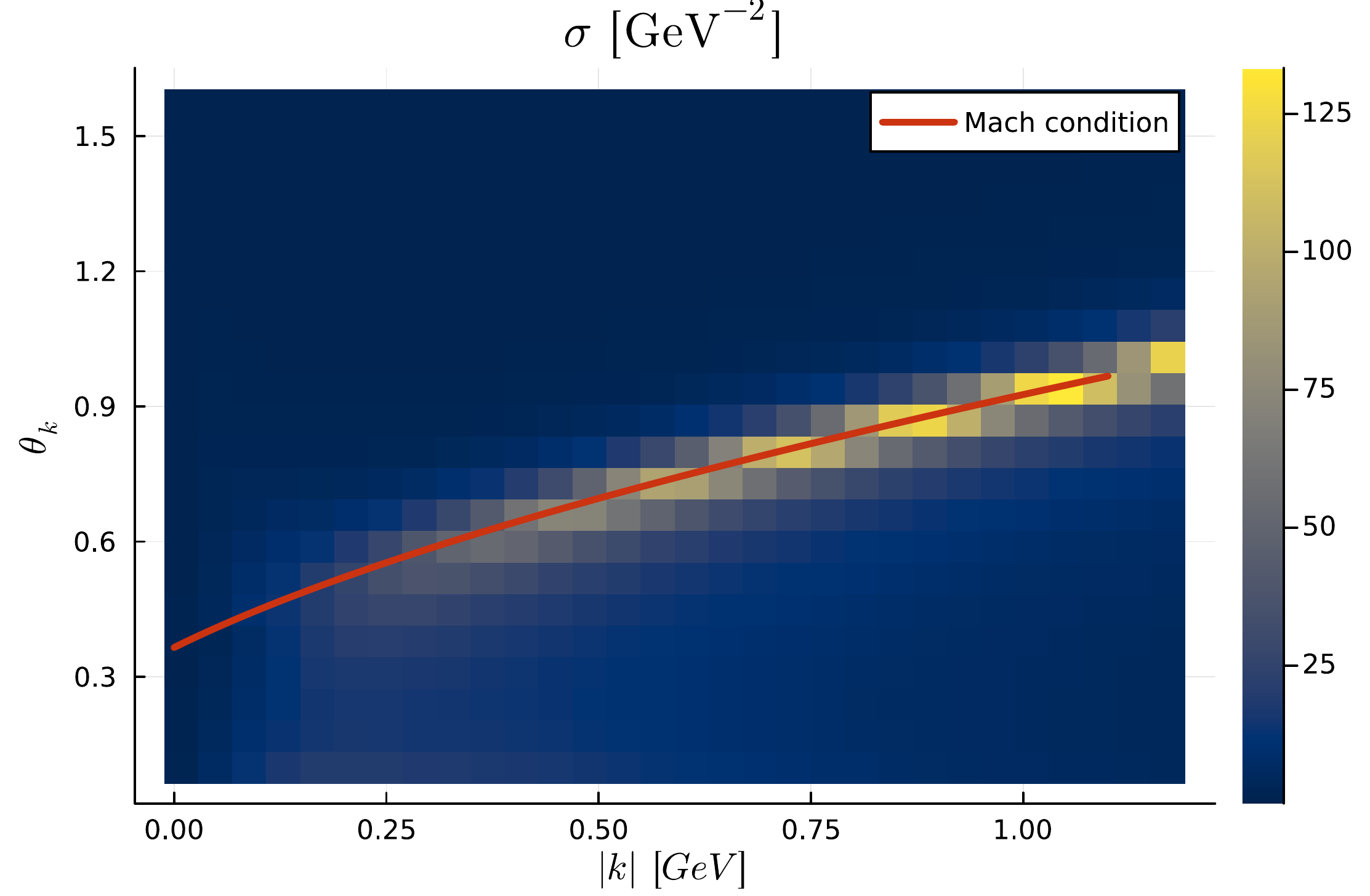}
\caption{}
\label{fig:xsec}
\end{subfigure}
\caption{(a) Leading-order heavy quark--phonon scattering: direct (top) and crossed (bottom) diagrams. (b) Cross section~\cite{Kirchner:2025ehq} vs.\ the incoming phonon momentum $|\vec k|$ and its angle $\theta_k$ to the heavy quark, for a charm quark with $|\vec p| = 1$~GeV, $T=0.3$~GeV, $c_s^2=1/3$; the solid curve is the absorption branch of Eq.~\eqref{eq:kin}.}
\label{fig:scattering}
\end{figure}

\section{Heavy quark--phonon scattering}

The width enters directly into heavy quark--phonon scattering, $Q(p)+\pi(k)\to Q(p')+\pi(k')$, which at leading order proceeds through the two diagrams of Fig.~\ref{fig:diagrams}, built from two vertices of Eq.~\eqref{eq:vertex}: the heavy quark absorbs the incoming phonon either before emitting the outgoing one or, in the crossed (Compton-like) diagram, after it. Between the two vertices the heavy quark can be on its mass shell, and the width computed above regulates the corresponding poles~\cite{Kirchner:2025ehq},
\begin{equation}
\frac{1}{2p\cdot k + k^2 + i\,m\,\Gamma(p+k)} \qquad\text{and}\qquad \frac{1}{-2p\cdot k' + k'^2 + i\,m\,\Gamma(p-k')} .
\label{eq:prop}
\end{equation}
The real part of the first denominator vanishes exactly when the incoming phonon can be absorbed on shell, i.e.\ on the absorption branch of Eq.~\eqref{eq:kin}. As a result, the cross section in Fig.~\ref{fig:xsec} peaks sharply, for each phonon momentum $|\vec k|$, at the angle $\theta_k$ given by this branch. For soft phonons, $|\vec k|\to 0$, it reduces to the Mach condition $\cos\theta_k = c_s/v$, and the second term is the recoil correction at finite phonon momentum. This is the angle at which a supersonic source emits sound, whose wavefronts form its Mach cone, so a supersonic heavy quark scatters preferentially with phonons moving in that direction. Both the temperature and the sound speed shape the resonance~\cite{Kirchner:2025ehq}: its height drops as $1/T$, while a smaller $c_s$ enlarges the heavy-quark width, which moves the peak to larger $\theta_k$ and broadens it. Scattering off quasiparticle partons has no such resonance.

\section{Conclusion and outlook}

The QHEFT describes the interaction of heavy quarks with the QGP systematically, through symmetry and power counting. We showed that the heavy-quark width grows steeply above the speed of sound, where phonon emission sets in, and that heavy quark--phonon scattering is resonantly enhanced along the Mach condition. These results show how the soft interaction of a heavy quark with the QGP can be computed in terms of the collective modes of the medium. Fixing the Wilson coefficients, above all $\tilde M'(1)$, by matching to holographic or lattice QCD or through comparison with data will set the size of these effects, and the scattering rates provide input for heavy-quark transport coefficients in a Langevin description.

\end{document}